# Optical Crossbars on Chip:
# a comparative study based on worst-case losses

Hui Li[1], Sébastien Le Beux[1]*, Gabriela Nicolescu[2], Jelena Trajkovic[3] and Ian O'Connor[1]

[1] Lyon Institute of Nanotechnology, INL-UMR5270 Ecole Centrale de Lyon, Ecully, F-69134, France

[2] Computer and Software Engineering Dept. Ecole Polytechnique de Montréal Montréal (QC), Canada

[3] Electrical and Computer Engineering Department Concordia University Montreal, QC, Canada

* Contact author: sebastien.le-beux@ec-lyon.fr

*Abstract* — The many cores design research community have shown high interest in optical crossbars on chip for more than a decade. Key properties of optical crossbars, namely a) contention free data routing b) low latency communication and c) potential for high bandwidth through the use of WDM, motivate several implementations of this type of interconnect. These implementations demonstrate very different scalability and power efficiency ability depending on three key design factors: a) the network topology, b) the considered layout and the c) the injection losses induced by the fabrication process. In this paper, the worst-case optical losses of crossbar implementations are compared according to the factors mentioned above. The comparison results has the potential to help many cores system designer to select the most appropriate crossbar implementation according, for instance, to the number of IP cores and the chip die size.

## I. INTRODUCTION

Technology scaling down to the ultra deep submicron domain provides for billions of transistors on chip, enabling the integration of hundreds of cores. Many core designs integrating interconnect that can support low latency and high data bandwidth are being increasingly required in modern embedded systems to address the increasing power and performance constraints of embedded applications. Designing such systems using traditional electrical interconnect represents a significant challenge: due to capacitive and inductive coupling [11], interconnect noise and propagation delay of global interconnect increase. The increase in propagation delay requires global interconnect to be clocked at a very low rate, which limits the achievable bandwidth and overall system performance.

In this context, Optical Network-on-Chip (ONoC) is an emerging technology considered as one of the key solutions for the future generation of on-chip interconnects. It relies on optical waveguides to carry optical signals, so as to replace electrical interconnect and provide the low latency and high bandwidth properties of the optical interconnect. Moreover, 3D integration technologies allow for both optical and electrical layers to be stacked. Proposals for ONoC can, thus, realistically envision the integration of sufficient photonic devices for fast chip-length communications [8][2][3].

Among the proposed ONoCs, the crossbar-based solutions gain considerable interest among the major players in the field. The efficient crossbar solutions rely on passive Microring Resonators (MRs), and they do not require any arbitration [1][4] due to the dedicated point-to-point connections between IP cores. In these networks, the signals propagation relies on Wavelength Division Multiplexing (WDM). Comparing the proposed crossbars is a tedious task, since it requires considering both network characteristics and technological data, assuming layout constraints.

In this paper, we compare proposed crossbars according to their worst-case losses, which is a key metric to evaluate the ONoC scalability and power efficiency. The worst-case losses can be estimated by considering the losses in the network (e.g. from waveguide crossing and waveguide length) and the optical losses value (e.g. propagation loss).

The paper is structured as follows. Section II presents the considered architecture model, topologies and implementations. Section III presents the loss model and a design methodology for a given crossbar ORNoC is given in Section IV. Section V gives the comparison results and Section VI concludes the paper.

## II. ARCHITECTURE MODEL, TOPOLOGIES AND IMPLEMENTATIONS

In this section, we describe the considered ONoC architecture model, the associated topologies and implementations.

### A. Architecture Model Overview

Figure 1 illustrates the considered 3D architecture model. It is composed of an electrical layer implementing NxN IP cores and an optical layer implementing ONoC. In our study, we assume N is an even number but the work could be easily extended for odd values and for NxM IP cores architectures. The optical network in the optical layer is composed of on-chip laser sources [9], MRs, and photodetectors. The ONoC is connected to the IP cores through Through Silicon Vias (TSV) [16]. Numerous ONoCs relying on WDM were proposed. Among these networks, wavelength routing scheme can be used to propagate data from a source IP core to a destination IP core, thus leading to a contention-free network (without need for arbitration), with high throughput, and low latency.

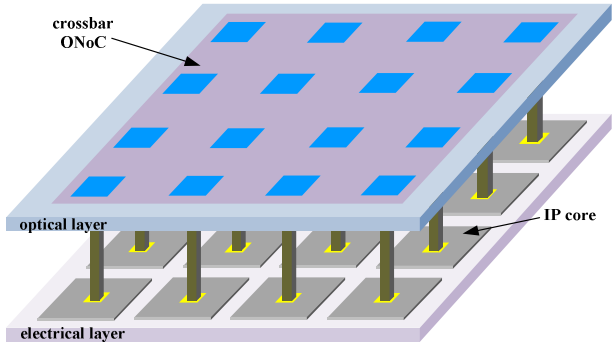

**Figure 1: The crossbar ONoC is implemented on the optical layer and it interconnects IP cores located on the electrical layer**

In this work, we compare ONoC architecture implementing crossbar functionality by considering the use of on-chip lasers. Indeed, efficient on-chip lasers usually require the inclusion of III–V semiconductors: gallium arsenide (GaAs) or indium phosphide (InP) are currently considered to be the best options. Microlasers, based on microdisk structures coupling light evanescently from the cavity resonant mode to the guided mode in an adjacent silicon waveguide, are sufficiently compact as to be implemented in a large number and at any position. For a given wavelength, the size of an on-chip laser is of the same order of magnitude as the size of an MR used to modulate continuous waves emitted by off chip lasers, which leads to a similar on-chip size for both approaches. While on-chip laser sources require the use of less mature technologies compared to their off-chip counterpart, they have the potential to provide the following three key advantages:

- Easier and more efficient integration by relaxing layout constraints: in case of on-chip lasers, it is not necessary to distribute the light from an external source to the modulators (e.g. through the so called *power waveguide* in Corona [7]). Relaxing such constraints contributes to reducing the number of waveguide crossings or even to avoiding them altogether in the ring topology.
- Higher scalability by keeping the architecture fully distributed, which is not achievable by considering centralized off-chip lasers.
- Lower power by reducing the worst-case communication distance. This corresponds to the distance from the source IP to the destination IP for on-chip laser based architectures, while for off-chip laser based architecture, this distance includes also the distance *from the off-chip laser to the source IP*. Shorter distance consequently reduces the optical losses and hence the minimum required laser output power. Moreover, the power consumption can be further improved by locally turning off the laser when no communication is required.

*B. Passive ONoC Architecture Implementations*

The crossbar network topologies considered in this study are 1) Matrix [17], 2) λ-router [1], 3) Snake [10], 4) ORNoC [4], as shown in Figure 4. In the figure, each column is dedicated to a topology and the lines give their i) graphical representation, ii) implementation characteristics, iii) layout and iv) loss model. This section briefly introduces these implementations and illustrates the way they can be used to interconnect 2x2 IP cores. We also illustrate the layout for 4x4 IP cores architecture, and evaluate the number of required optical devices, assuming N is an even number.

*1) Matrix*

The first crossbar topology relies on a traditional Matrix-like structure. Figure 2 (a) illustrates a simple example where 4 IP cores are interconnected using the Matrix. Full connectivity is considered, which leads to a total of 16 MRs in this example. By considering only inter-IP communications, one MR per line of the Matrix can be removed. For NxN IP cores, $(N^2-1) \times N^2$ passive MRs are, thus, used to implement the crossbar itself. The transmitters are composed of on-chip laser sources, and the receivers are composed of photodetectors and passive MRs that drop the signal onto the photodetector (not illustrated in the figure for sake of clarity). Because we focus on crossbar networks, we assume dedicated communication between all the IP cores through spatial WDM. As a consequence $(N^2-1) \times N^2$ laser sources, photodetectors and passive MR are required in the network interface. It is worthwhile noting that all topologies considered in this paper require the same. Matrix topology uses $N^2-1$ wavelengths to implement all the communications.

Figure 2 also presents two possible layouts $layout_A$ and $layout_B$, which are designed to A) avoid any waveguide crossing between the network interfaces and the crossbar itself and B) reduce the worst-case distance between IP cores. The crossbar is located in the middle of the optical layer for layout optimization purposes (it is represented as a box for the sake of clarity). It interconnects 16 inputs (in red lines) with 16 outputs (in black lines) through 240 MRs. The same layouts will be assumed to interface with λ-router and Snake networks.

*2) λ-router*

λ-router is a multistage network topology relying on WDM to propagate optical signals from input to output ports. Compared to the Matrix, the multistage structure allows reducing the number of waveguide crossing in the worst-case scenario (6 and 3, for Matrix and λ-router, respectively, in the case of 4 IP cores). This is achieved by assuming symmetric 2x2 switches structure relying on 2 identical MRs. The initial structure of λ-router would assume 240 MRs, for the architecture with 4x4 IP cores, but a reduction method [1] reduces the network complexity by managing only the required optical connections and by removing the unused MRs. As a result, 224 MRs are required to implement the network.

*3) Snake*

Snake is, also, a multistage network topology. It has the same properties as the λ-router. The only difference is the distribution of the MRs in the network, which leads to the more compact layout compared to λ-router, with the side effect of different waveguide length between different input and output pairs. Similarly to λ-router, a reduction method adapted from [1] can be applied to remove unused MRs.

*4) ORNoC (Optical Ring Network-on-Chip)*

In ORNoC, each IP core communicates with another IP through waveguides forming a ring. The following operations are performed:

- Injection: the IP core injects an optical signal into a waveguide through its output port data. The wavelength of the signal specifies the destination of the IP core;

- Pass through: the incoming signal propagates along the waveguide (i.e. no MR with the same resonant wavelength is located along the waveguide);
- Ejection: the incoming optical signal is ejected from the waveguide and is redirected to the destination IP core. This is achieved by an MR located along the waveguide and with the same resonant wavelength as the signal.

In ORNoC, the same wavelength can be used to realize multiple communications on the same waveguide, at the same time. Furthermore, and multiple waveguides can be used to interface IP cores. Both clockwise (C) and counter-clockwise (CC) rotation can be considered for signal propagation, where each direction is realized on a separate waveguide. For this comparative study, we consider two versions of ORNoC: $ORNoC_C$ and $ORNoC_{C-CC}$ that rely on: only the C rotation, and both C and CC rotations, respectively. Both are illustrated in Figure 2: blue and red lines represent C and CC directions separately. Compared to the other networks, no MR is used in the network itself, (i.e. MRs are used only in the network interfaces. This leads to a reduction of the total number of optical devices. However, the ring structure implies crossing intermediate interfaces. This leads to an increase of the minimum number of wavelengths to be used (6 and 3 wavelengths are required to interconnect 4 IP cores with $ORNoC_C$ and $ORNoC_{C-CC}$, respectively). If the maximum number of wavelengths in the network is reached, then waveguides can be added in order to realize all the communication, without any impact on the layout complexity and the waveguide crossing, by considering a serpentine layout.

| | | 1) Matrix | 2) λ-router | 3) Snake | 4a) $ORNoC_C$ | 4b) $ORNoC_{C-CC}$ |
|---|---|---|---|---|---|---|
| Structural view | | 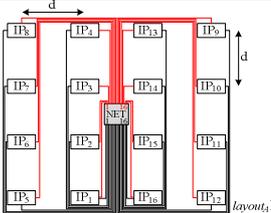 | | | | |
| No. of resources | $MR_{net}$ | $N^4$ | $(N^2-1) \times N^2$ | $(N^2-1) \times N^2$ | 0 | 0 |
| | $MR_{less}$ | $(N^2-1) \times N^2$ | $(N^2-2) \times N^2$ | $(N^2-2) \times N^2$ | 0 | 0 |
| | $MR_{det}$ | $(N^2-1) \times N^2$ | | | | |
| | $NB_{Laser}$ | $(N^2-1) \times N^2$ | | | | |
| | $NB_{wl}$ | $N^2-1$ | $N^2$ | $N^2$ | $(N^2-1) \times N^2 / 2$ | $(N^2-1) \times N^2 / 4$ |
| Layout | | 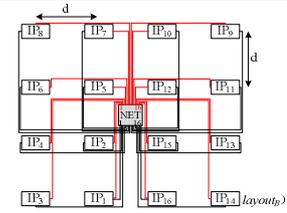 | | | | |
| Loss model | $D_{max}$ | $Layout_A: \begin{cases} 3d & N: 2 \\ 4 \times \lfloor (N-1)/2 \rfloor \times d + 2 \times N \times d & N: \text{even and } N>2 \end{cases}$ $Layout_B: \quad 2 \times (N-1) \times d$ | | | $(N^2-2) \times d + (N-1) \times d$ | $(\lfloor N^2/2 \rfloor -1) \times d + (N-1) \times d$ |
| | $N_{crossing}$ (network and layout) | $2 \times N^2 - 3$ | $N^2 - 1$ | $2 \times N^2 - 5$ | 0 | 0 |
| | | $Layout_A: 0$ | | | | |
| | | $Layout_B: 3 \times N^2/4 + N/2 - 1$ | | | | |
| | $N_{drop}$ | 2 | 2 | 2 | 1 | 1 |

**Figure 2: Summary of considered ONoC: 1) Matrix, 2) λ-router, 3) Snake, 4a) $ORNoC_C$ and 4b) $ORNoC_{C-CC}$**

## III. OPTICAL LOSS MODEL

This section presents the proposed optical loss models for crossbar comparison.

The worst-case losses in the optical path $L_{wc}$ for each network is defined by assuming NxN IP cores and by assuming N to be an even number greater than 2. We assume on-chip laser sources for all topologies, which contribute to reduced number of waveguide crossing compared to the off-chip laser counterpart. The loss model is given by the expression (1). $L_{wc}$ depends on: the total propagation loss in the waveguide $L_{waveguide}$, total loss due to waveguide crossing $L_{crossing}$, and drop loss $L_{drop}$ when a signal encounters a MR with the same wavelength, all given in dBs. We assume a negligible bending loss:

$$L_{wc}^{dB} = L_{waveguide}^{dB} + L_{cros\sin g}^{dB} + L_{drop}^{dB} \quad (1)$$

- $L_{waveguide} = P_{propagation} \times d_{max}$, with $P_{propagation}$ (in dB/cm) the intrinsic propagation loss of the optical signal in the waveguide and $d_{max}$ (in cm) the longest distance between the source and destination. It depends on the layout represented in Figure 2. A key metric to define $d_{max}$ is the distance $d$ between two neighboring interfaces to the IP cores;
- $L_{cros\sin g} = P_{cros\sin g} \times N_{cros\sin g}$;
- $L_{drop} = P_{drop} \times N_{drop}$;

with $P_{crossing}$ and $P_{drop}$ the injection loss occurring in waveguide crossing and drop operation, respectively, and $N_{crossing}$ and $N_{drop}$ their respective number of occurrences in the worst case scenario.

Considering both technological and structural values related to the fabrication process and the network topology enables a fair comparison between networks.

Figure 2 gives the values $D_{max}$, $N_{crossing}$ and $N_{drop}$ for the considered networks. For Matrix, λ-router and Snake networks, both layouts are considered, which leads to bigger longest distance in case A and more additional waveguide crossing in case B. We do not consider the distance between inputs and outputs of the network itself. Two drop operations occur, one in the network itself and one more in the receiver interface to drop the signal into the photodetector.

Both ORNoC$_C$ and ORNoC$_{C-CC}$ do not suffer from any waveguide crossing and the signal is dropped only once in the receiver part ($N_{crossing}=0$ and $N_{drop}=1$). However, the considered serpentine layout implies that $d_{max}$ increases more rapidly when compared to the other networks. It is worth noticing that $d_{max}$ is significantly reduced for the C-CC case compared to the C case. This will result in a lower worst-case loss, which directly contributes to the energy-efficiency of ORNoC$_{C-CC}$. The following section gives the design methodology assumed for ORNoC in order to reduce both the number of required wavelengths and the worst case distance between a source IP and a destination IP.

## IV. ORNoC WAVELENGTH ASSIGNMENT METHODOLOGY IN ORNoC

The efficient design of ORNoC requires careful wavelength assignment between IP cores to reduce the number of wavelengths. This section details the methodology for ORNoC$_C$ and ORNoC$_{C-CC}$.

In ORNoC$_C$, a single direction is used to propagate the signals. Therefore, in the the connectivity matrix, we allocate the same wavelength i) between IP$_i$ and IP$_j$ and ii) between IP$_j$ and IP$_i$. Following this method, one wavelength is used in the whole ring to implement 2 connections, thus leading to an efficient assignment of the wavelength (i.e. there is no ring segment unoccupied by the wavelength). By considering the use of a single waveguide, the total number of required wavelengths is equal to half of the total number of connections in the network, i.e. $N_{wl}=(N^2-1)xN^2/2$. By considering the use of multiple waveguides, the total number of used wavelengths can be reduced by reusing a same set of wavelengths in different waveguides. Figure 3 a) illustrates an example of wavelength assignment between a source (S) and a destination (D) for the design of a crossbar connecting 4 IP cores. In this example, 6 wavelengths are required ($\lambda_0...\lambda_5$), which is twice the number of required wavelengths in Snake, λ-router and Matrix. The large number of wavelengths is mainly due to the fact that each wavelength can be used only twice per waveguide.

| S\D | IP$_1$ | IP$_2$ | IP$_3$ | IP$_4$ | | S\D | IP$_1$ | IP$_2$ | IP$_3$ | IP$_4$ |
|---|---|---|---|---|---|---|---|---|---|---|
| IP$_1$ | - | $\lambda_0$ | $\lambda_1$ | $\lambda_2$ | | IP$_1$ | - | $\lambda_2$ | $\lambda_0$ | $\lambda_1$ |
| IP$_2$ | $\lambda_0$ | - | $\lambda_3$ | $\lambda_4$ | | IP$_2$ | $\lambda_2$ | - | $\lambda_2$ | $\lambda_1$ |
| IP$_3$ | $\lambda_1$ | $\lambda_3$ | - | $\lambda_5$ | | IP$_3$ | $\lambda_0$ | $\lambda_2$ | - | $\lambda_0$ |
| IP$_4$ | $\lambda_2$ | $\lambda_4$ | $\lambda_5$ | - | | IP$_4$ | $\lambda_1$ | $\lambda_1$ | $\lambda_0$ | - |

**Figure 3: Connectivity matrix for a) ORNoC$_C$ and b) ORNoC$_{C-CC}$**

In ORNoC$_{C-CC}$, the long distance communication can be avoided by implementing the connection through a shortest path assignment schemes, relying on the selection of the appropriate direction on a different ring. Communications from IP$_i$ to IP$_j$ and from IP$_j$ to IP$_i$ are realized through 2 separated waveguides propagating signals in opposite direction. For sake of regularity and symmetry in the connectivity matrix, the same wavelength is used for bidirectional communications. The wavelengths are assigned as follow: starting from source core IP$_X$, first a wavelength is assigned to the longest distance communication in direction C, to a destination core IP$_Y$; second, the same wavelength is assigned to communication from IP$_Y$ to the longest distance communication (still in direction C), to a destination core IP$_Z$. We apply the same assignment to the following longest distance communication until source core IP$_X$ is reached, meaning that the wavelength is used on the whole ring (i.e. the wavelength is efficiently used on the ring). The same process is applied starting from each IP core with a different wavelength. The same algorithm iterates with other wavelengths until a wavelength is assigned to each connection. Figure 3 b) illustrates the connectivity matrix for the 4 IP cores architecture example. Blue and red colors indicate the use of C and CC direction for signal propagation, respectively. Three wavelengths are required when considering the use of a single waveguide for each direction (e.g. by considering 3 waveguides, a single wavelength would be required). In this example, each wavelength is used only up to twice on a single waveguide, which is due to the small number of IP cores; wavelengths can be further reused as the number of IP cores increases, which contributes to the improvement of the communication density.

## V. COMPARATIVE STUDY

We compare the presented implementations according to the worst-case losses. In a first comparison, all the networks are considered by assuming a given set of technological value extracted from Table 1. In a second comparison, we further compare the networks assuming various design parameters.

**Table 1: Injection Loss parameters**

| Optical loss | $P_{crossing}$ | $P_{propagation}$ | $P_{drop}$ |
|---|---|---|---|
| Pan (2010) [12] | 0.05 | 1 | 1.5 |
| Kirman (2010) [13] | 0.12 | 1 | 1 |
| Biberman (2011) [6] | 0.05 | 0.5 | 0.5 |
| Koka (2012) [14] | 0.2 | 0.1 | 1.5 |

### A. Worst-case losses evaluation

We first assume a fixed 4cm² die size and evaluate the losses for different architecture size: N=2, 4, 6 and 8; where the distance between IP cores decreases as the number of IP cores increases, d=10mm, 5mm, 3.33mm and 2.5mm, respectively. Figure 4 a) and b) illustrate the evaluation results for injection losses with the parameter values given by Pan and Biberman, respectively. If we compare different layouts for Matrix, λ-router and Snake topologies, using the values from Pan, (Figure 4 a), $layout_B$ uniformly outperforms $layout_A$, for all network sizes. By considering the set of values from Pan (Figure 4 a), $layout_B$ outperforms $layout_A$ for Matrix, λ-router and Snake networks. By considering values from Biberman (Figure 4 b), the same conclusion can be made for architectures containing up to 6x6 IP cores. However, for 8x8 IP cores, worst case loss is lower for $layout_A$, due to the lower propagation loss in the waveguide, thus highlighting the better scalability of this layout. It is worth noticing that other layouts could provide good tradeoff. ORNoC$_{C-CC}$ is the most scalable network despite the long distance introduced by the serpentine layout. By considering values from Biberman, ORNoC$_{C-CC}$ is the most scalable network, with 5.25dB in the worst case path for 8x8 IP cores, followed by λ-router using $layout_B$ (so called λ-router-b in the figure) with 8.45dB, thus achieving 37.9% improvement. By assuming parameters from Koka (not illustrated), the worst-case loss in ORNoC$_{C-CC}$ and λ-router-b become 2.45dB and 26.15dB, thus leading to a 90.6% improvement for ORNoC$_{C-CC}$. Because of the rather large distance implied by the die size we consider, ORNoC$_C$ does not exhibit a good scalability.

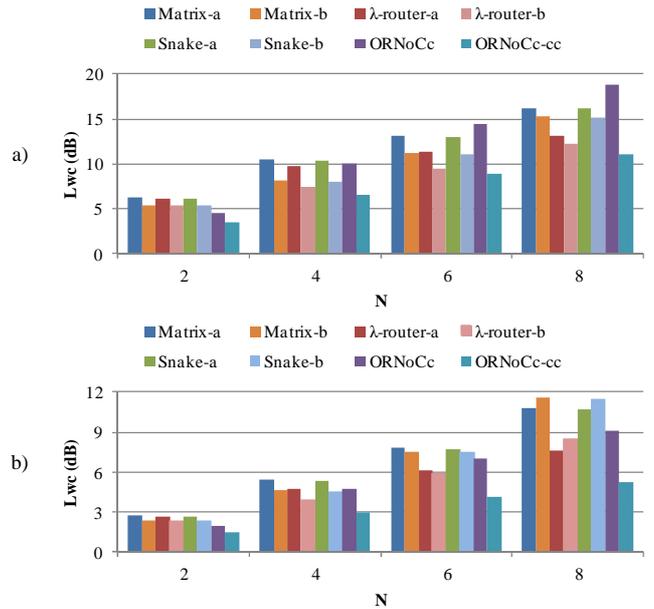

**Figure 4: Worst-case losses evaluation for various number of IP cores assuming injection loss parameters from a) Pan [12] and b) Biberman [6]**

For a 8x8 architecture with a single waveguide per direction, ORNoC$_{C-CC}$ requires 1008 wavelengths compared to 63 wavelengths for Matrix and 64 for Snake and λ-router. 1008 is not a realistic value for the number of wavelengths; however, it is important to notice that additional waveguides can be used in ORNoC to satisfy the constraints on the maximum number of wavelengths, which can be achieved without any waveguide crossing because of the 3D architecture and the use of on-chip laser sources. Following the methodology from [4], ORNoC would require 16 waveguides if we consider an optimistic maximum number of 64 wavelengths per waveguides, and 63 waveguides if we consider more realistic scenario with 16 wavelengths per waveguide. If such constraints on the number of wavelengths must be respected for Matrix, Snake or λ-router, this can be achieved by considering the use of multiple networks, which implies additional waveguide crossing [15]. With ORNoC$_{C-CC}$, no waveguide crossing is required, the layout is regular and $d_{max}$ is reduced compared to ORNoC$_C$, which make the network implicitly scalable without any custom place-and-route tool [10][5].

Figure 5 represents the worst case loss assuming parameters given by Biberman, size of 8x8 IP cores, and various distance between IP cores (d=0.125, 0.25, 0.5, 1 and 2mm). The impact of the distance increase is the greatest for ORNoC$_C$ and ORNoC$_{C-CC}$, due to the serpentine layout. Still, even for a 2mm distance, which implies a realistic 2.56cm² die size, ORNoC$_{C-CC}$ remains the most power efficient network.

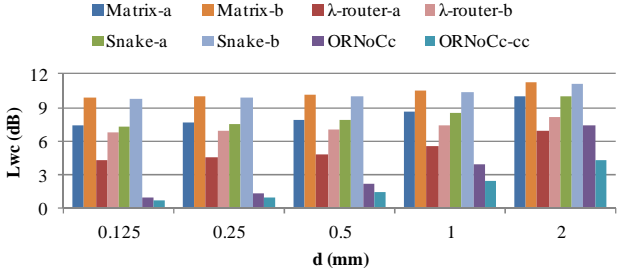

**Figure 5: Evaluation of the impact of the distance between IP cores on the worst case losses**

*B. Implementation Comparison*

In order to further explore the design space, for the example of 8x8 IP cores, and various distances, we consider a range of 0-2dB for propagation losses and a range of 0-0.2dB for waveguide crossing loss.

Figure 6 illustrates comparison results for the implementation of λ-router according to *layout$_A$* (i.e. without waveguide crossing) and *layout$_B$* (reduced waveguide length), assuming $P_{drop}$=1dB. We also plot the values from Table 1. The area below each line represents the design space for which the worst case loss is lower for *layout$_A$*; the area above the line gives the design space where the worst case loss is lower for *layout$_B$*, and the line itself represents the designs with the same worst case losses for both layouts. This further helps determine the most appropriate layout, for a given set of injection loss values, and a given distance between IP cores.

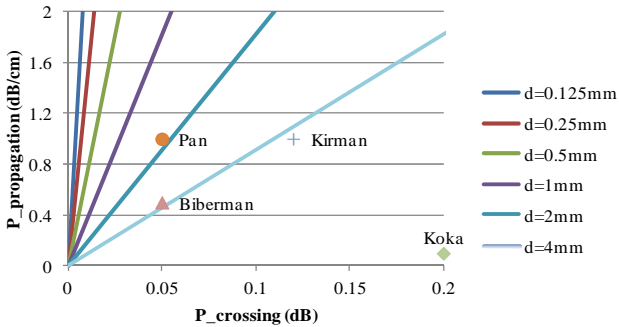

**Figure 6: Comparison of *layout$_A$* and *layout$_B$* for the implementation of λ-router (8x8 IP cores, $P_{drop}$=1dB)**

Figure 7 further compares Snake and ORNoC$_{C-CC}$: the area below a line represents the design space for which ORNoC$_{C-CC}$ provides lower worst-case losses, thus highlighting the advantage of ORNoC$_{C-CC}$ compared to Snake.

These comparisons highlight the importance of technological parameters, layout and network characteristics to evaluate the worst-case optical loss. For a given set of technological value (e.g. crossing loss), certain topology and layout may be more advantageous, which may significantly impact the overall power efficiency of the crossbar.

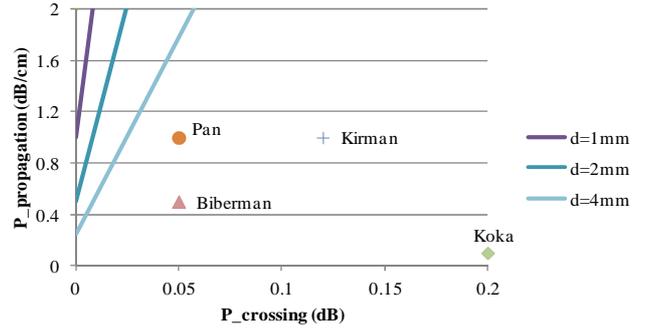

**Figure 7: Comparison between Snake vs ORNoC$_{-CC}$ (8x8 IP cores, $P_{drop}$=1dB)**

## VI. CONCLUSION

Optical crossbars on chip represent an efficient interconnect solution for many cores architectures. Various crossbar implementations are possible and their worst-case losses rely on topological, physical and technological aspects. In this paper, we compare possible crossbar implementations relying on matrix, multistage and ring-based networks. For a given number of IP cores to interconnect and a given die size, our approach allows to identify the implementation characterized by the lower worst-case optical losses, i.e. the most power efficient solution. For the explored design space, ring-based topology implementations exhibit higher power efficiency compared to matrix based and multistage-based network implementations. The approach was applied to passive and fully interconnected networks but it can be extended to active networks requiring resources allocation mechanism. We will focus on this aspect in our future work.

### ACKNOWLEDGMENT

Sébastien Le Beux is supported by a Région Rhône-Alpes grant.